\documentclass[reprint,amsmath,amssymb,aps,prb]{revtex4-1}
\usepackage{blindtext}
\usepackage{graphicx}
\usepackage{dcolumn}
\usepackage{bm}
\usepackage[mathlines]{lineno}
\usepackage[english]{babel}
\usepackage{xcolor}

\usepackage{multirow}

\usepackage{eqnarray,amsmath}
\usepackage{adjustbox}
\usepackage{booktabs}
\usepackage[T1]{fontenc}
\usepackage[utf8]{inputenc}
\usepackage{lipsum}
\usepackage[colorlinks]{hyperref}
\usepackage{xspace}
\usepackage[toc,page]{appendix}

\usepackage{placeins}
\usepackage{float}

\newcommand\T{\rule{0pt}{3ex}}       
\newcommand\B{\rule[-0.9ex]{0pt}{0pt}} 
\newcommand{\TB}{\T\B} 

\begin{document}
\title{A deep investigation of NiO and MnO through the first principle calculations and Monte Carlo simulations} 
\author{Mojtaba Alaei}
\email{m.alaei@iut.ac.ir}
\affiliation{Department of Physics, Isfahan University of Technology, Isfahan 84156-83111, Iran } 
\author{Homa Karimi} 
\email{homa.karimi@alumni.iut.ac.ir}
\affiliation{Department of Physics, Isfahan University of Technology, Isfahan 84156-83111, Iran } 
\date{\today}
\begin{abstract}
In this study, we use Hubbard-Corrected density functional theory (DFT+$U$) to derive spin model Hamiltonians consisting of Heisenberg exchange interactions up to 
the fourth nearest neighbors and bi-quadratic interactions. 
We map the DFT+$U$ results of several magnetic configurations to the Heisenberg spin model Hamiltonian to estimate Heisenberg exchanges. 
We demonstrate that the number of magnetic configurations should be at least twice the number of exchange parameters to estimate exchange parameters correctly. 
To calculate biquadratic interaction, we propose specific non-collinear magnetic configurations that do not change the energy of the Heisenberg spin model.
We use classical Monte Carlo (MC) simulations to evaluate DFT+$U$ results. 
We obtain the temperature dependence of magnetic susceptibility and specific heat to determine the Curie-Weiss and N\'eel temperatures. 
The MC simulations reveal that although the biquadratic interaction can not change the N\'eel temperature, it modifies the order parameter.
We indicate that for a fair comparison between classical MC simulations and experiments, 
we need to consider the quantum effect by applying $(S+1)/S$ correction in classical MC simulations. 
\end{abstract}
\maketitle
\section{\label{Introduction}Introduction}
\begin{figure}
    \centering
      \includegraphics[width=0.80\columnwidth]{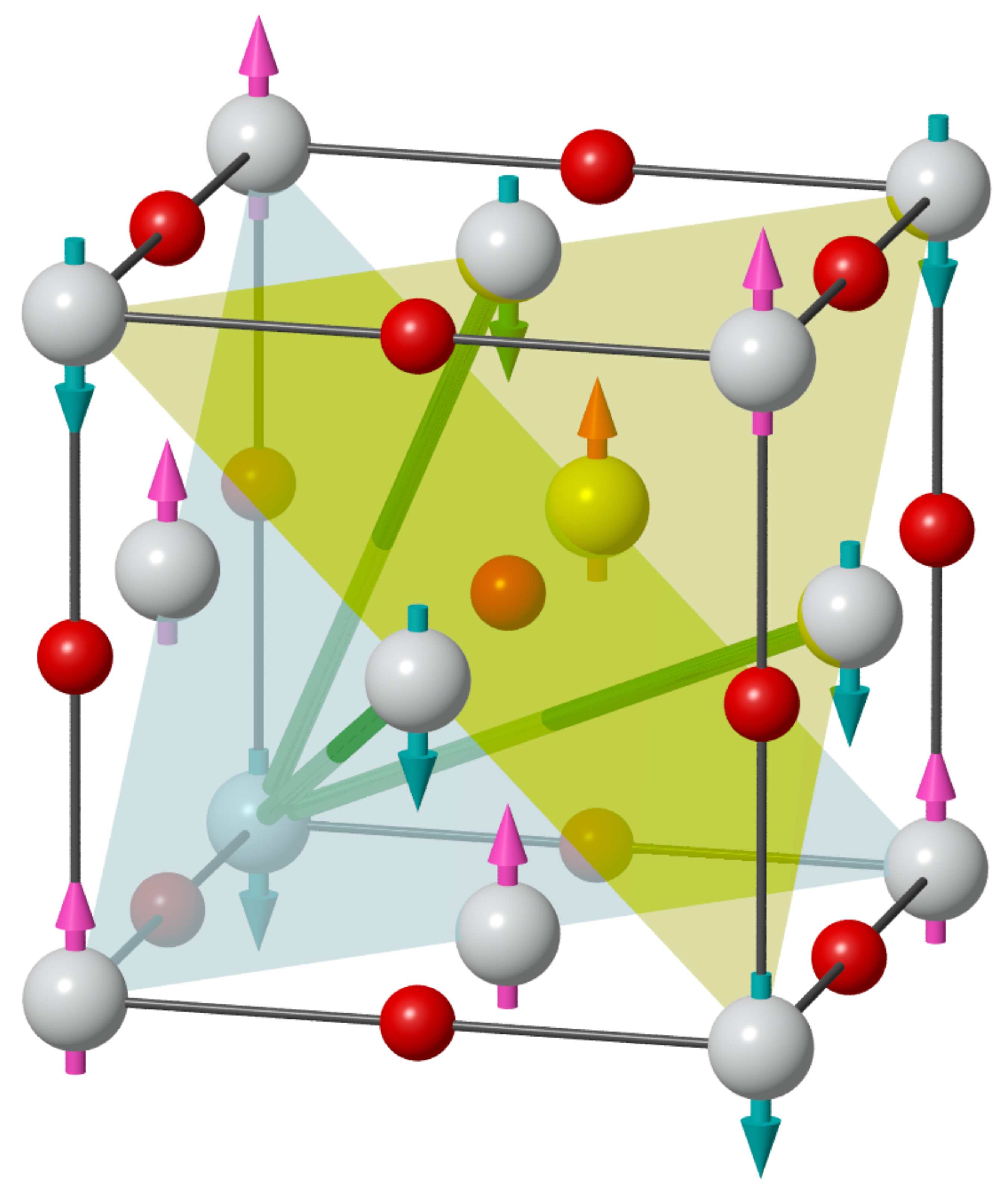}
      \caption {(Color online) The cubic and AF primitive cells of NiO and MnO are indicated by thin black cylinders and thick green cylinders, respectively. 
                               The big gray spheres represent transition metal atoms (Mn, Ni), and the small red spheres indicate oxygen atoms. 
                               The blue and yellow planes show two successive  (111) planes. The magnetic moments are arranged ferromagnetically in each plane, 
                               while the order of magnetic moments between two successive planes is AF.}

    \label{fig:cubic}
\end{figure}

The magnetic and electronic structure properties of transition metal oxides (TMOs) have been widely investigated experimentally and theoretically. 
Since TMOs belong to highly correlated systems, they are ideal for challenging theoretical methods. 
MnO and NiO are two outstanding representations of the first row TMOs. 
Their magnetic structures have been thoroughly characterized by the neutron diffraction method \cite{Shull1949, Shull1951, Roth1,MnO2006, NiO2016}. 
Above their N\'{e}el temperatures\cite{Balagurov} 
(${\mathrm{T}_{N}\sim119\pm1 \mathrm{K}}$ for MnO and ${\mathrm{T}_{N}\sim523\pm2 \mathrm{K}}$ for NiO), 
both adopt the paramagnetic phase and rock salt structure. 
Their magnetic ions arrange in an antiferromagnetic (AF) type II ordering 
below the respective N\'{e}el temperatures. 
In the AF(II) phase, there are ferromagnetic orders inside ${\left(111\right)}$ planes, 
while two successive ${\left(111\right)}$ planes have the AF order (Fig.~\ref{fig:cubic}).

Since we need magnetic interaction between magnetic moments to get insight into thermodynamic properties, 
one of the main proposes of ab initio methods is to derive exchange interactions. 
Unfortunately, techniques such as dynamical mean-field theory (DMFT) demand many computational resources. 
Hence, we need methods based on density functional theory (DFT) for fast predictions. 
Regarding local density approximation (LDA) \cite{Kohn65,Rajagopal}  
and generalized gradient approximation (GGA) \cite{Langreth}, 
DFT can not comprehensively study the electronic structure of strongly correlated electron systems. 
That is because inherent self-interaction errors cause the inadequacy of exchange-correlation (XC) functional at localizing the valence d electrons, 
which dramatically affects the electronic prediction of ground-state, such as underestimating bandgap in TMOs. 
Therefore, DFT methods have been manipulated to overcome these deficiencies and include, but are not limited to, 
DFT+$U$ (LDA+$U$, GGA+$U$) \cite{ldau1,ldau2}, self-interaction correction (SIC) \cite{Szotek1993, Svane1990}, and hybrid functional \cite{Becke}.

Verifying microscopic magnetic interactions (such as exchange interactions)  
derived from DFT methods in strongly correlated oxides 
is a critical step in investigating the competence of the DFT methods. 
A recommended way is to compare the theoretical and experimental spin-wave dispersions. 
Another way is to compare the thermodynamical features, such as transition and Curie-Weiss temperature,
with experimental results. We can estimate these temperatures 
using Monte Carlo (MC) simulations of spin models containing exchange interactions.

This work attempts to derive Heisenberg and Biquadratic interaction through the DFT+$U$ method for MnO and NiO. 
One prevalent way to derive the Heisenberg exchange parameters using DFT+$U$ 
is to calculate the total energy differences of several (collinear) magnetic configurations. 
In this study, we demonstrate  that the number of magnetic configurations 
can drastically affect the estimation of Heisenberg exchange interactions. 
Contrary to most studies, we indicate that the number of magnetic configurations 
for such estimation should be at least twice of exchange parameters.
In addition, we introduce a simple method to derive biquadratic exchange interaction for TMOs, 
which has already been used for pyrochlore crystals~\cite{fef3-1}. 
After deriving exchange parameters, we use these parameters in classical MC simulations 
to obtain transition and Curie-Weiss temperatures. 
In this regard, we discuss the correct comparison of classical MC and experimental results.

The paper organizes as follows; we dedicate Section~\ref{Computational} to the details of computational methods. 
Section~\ref{Results} presents the exchange parameters calculated using DFT+$U$ and the technical points required
to obtain these parameters correctly. 
At the end of this section, we provide the MC results. In the final section, we present our conclusions.

\section{\label{Computational}Computational Details}
For ab initio DFT calculation, 
we employ full potential linear augmented plane wave (FL-LAPW) using Fleur code~\cite{fleur}. 
We set 2.0,  2.0, and 1.75 a.u. Muffin-tin radius  for Mn, Ni, and O, respectively. 
We use $k_{\mathrm max} = 4.5~ \mathrm{a.u.}^{-1}$ as  the plane wave expansion cutoff in the interstitial region.
We use the cubic cell containing eight atoms to calculate the biquadratic exchange parameter. 
In contrast, we construct a $2\times2\times2$ super-cell (containing 32 atoms) from the AF primitive cell (containing four atoms) 
for Heisenberg exchange parameters. 
In Fig. \ref{fig:cubic}, we illustrate the cubic and the AF primitive cell.
We set $8\times8\times8$ k-point mesh for the cubic cell and $4\times4\times4$ for the super-cell.       
We approximate electron-electron exchange-correlation energy using LDA with Perdew-Zunger (PZ) functional~\cite{pz} and 
GGA with Perdew-Burke-Ernzerhof  (PBE) functional~\cite{pbe}.
To correct the electron-electron interaction in d orbitals, we employ the DFT+$U$ approximation~\cite{ldau1,ldau2}.
There are two main approaches for calculating Hubbard $U$ and Hund exchange $J_{\mathrm{H}}$ from the first principles to use in DFT+$U$ calculations: 
Constrained density functional theory (cDFT)~\cite{cDFT1,cDFT2} and constrained random-phase approximation (cRPA)~\cite{cRPA1, cRPA2}. 
These methods have obtained the Hubbard $U$ parameter between 6.9 to 5.25 eV for MnO and 8.0 to 5.77 eV for NiO~\cite{ldau1, Uparameter1, Uparameter2}.  
Since our calculation indicates that using $U=6.9$ eV for MnO and $U=8.0$ eV results in smaller exchange Heisenberg interactions 
than the experiment, we examine $U$ from 6.9 to 4 eV for MnO and 8.0 to 5 for NiO to find proper Heisenberg exchanges. 
For the Hund exchange $J_{\mathrm{H}}$, we set 0.86 and 0.95  eV for MnO and NiO, respectively.

To obtain the spin model for MnO and NiO, we map DFT+$U$ results into the following spin model Hamiltonian:
\begin{equation}
H_{\mathrm{model}}=-\frac{1}{2} \sum_{i,j} J_{i,j} \mathbf{\hat{S}}_i \cdot \mathbf{\hat{S}}_j + B \sum_{<i,j>} (\mathbf{\hat{S}}_i\cdot \mathbf{\hat{S}}_j)^2,
\end{equation}
where $\mathbf{\hat{S}}_i$ indicates the magnetic moment direction of the $i$th site of transition metal atoms. 
The $J_{i,j}$, and $B$ indicate the strength of Heisenberg exchange and bi-quadratic exchange interaction~\cite{bq1,bq2,bq3} between the $i$th and $j$th  sites, respectively.
The summation notation $<i,j>$ for bi-quadratic interactions indicates that sites $i$ and $j$ are nearest neighbors, and each pair is included once.
Due to spin-orbit coupling, magnetic anisotropy (MA), and Dzyaloshinskii-Moriya (DM) are two typical additional interactions in magnetic systems. 
However, the MA is negligible for MnO and NiO~\cite{MAI}, and due to Moriya's rules~\cite{DMI}, 
the DM interaction does not exist for MnO and NiO.
We obtain the $J_{i,j}$ exchanges by mapping the DFT total energies of different magnetic configurations into the Heisenberg Hamiltonian model~\cite{Jij}.
For mapping DFT results into the Heisenberg Hamiltonian model, 
We calculate the DFT+$U$ total energy of 32 colinear magnetic configurations, 
then use the least-squares technique to estimate $J_{i,j}$ exchanges.
It is worth mentioning that in the collinear magnetic configurations, 
the bi-quadratic term of the Hamiltonian is degenerate and has no effect on the estimation of the Heisenberg exchange parameters.
To obtain the bi-quadratic interaction, we consider some non-collinear magnetic configurations 
in which the Heisenberg part of the Hamiltonian is degenerate.
We explain it in more detail in section~\ref{Results}.

To obtain temperature-dependent magnetic properties of MnO and NiO, 
such as the N\'eel temperature (${T_{\mathrm{N}}}$) and Curie-Weiss temperature (${\Theta_{\mathrm{CW}}}$), 
we employ the ESpinS package~\cite{espins} for MC simulations. 
We consider the replica-exchange method~\cite{ptmc1,ptmc2,ptmc3} (also known as the parallel tempering method) to
perform the classical MC simulations. 
%
%
%
We set the number of lattice sites around 6000-7000 (for MnO and NiO, we use a lattice size of $N=2\times15^3$). 
For each replica, at each temperature, the MC simulation was carried out with ${5\times10^{5}}$ sweeps of the lattice 
for equilibration and sampling separately. Measurements were done after each 5 MC sweeps to reduce the correlation between the data. 
In the PT algorithm, the swapping of replicas at the adjacent temperatures was allowed after each 10 MC sweeps.
\section{\label{Results}Results and Discussion}
\subsection{ Heisenberg coupling parameters from DFT calcualtions}
\begin{figure}
    \centering
      \includegraphics[width=0.95\columnwidth]{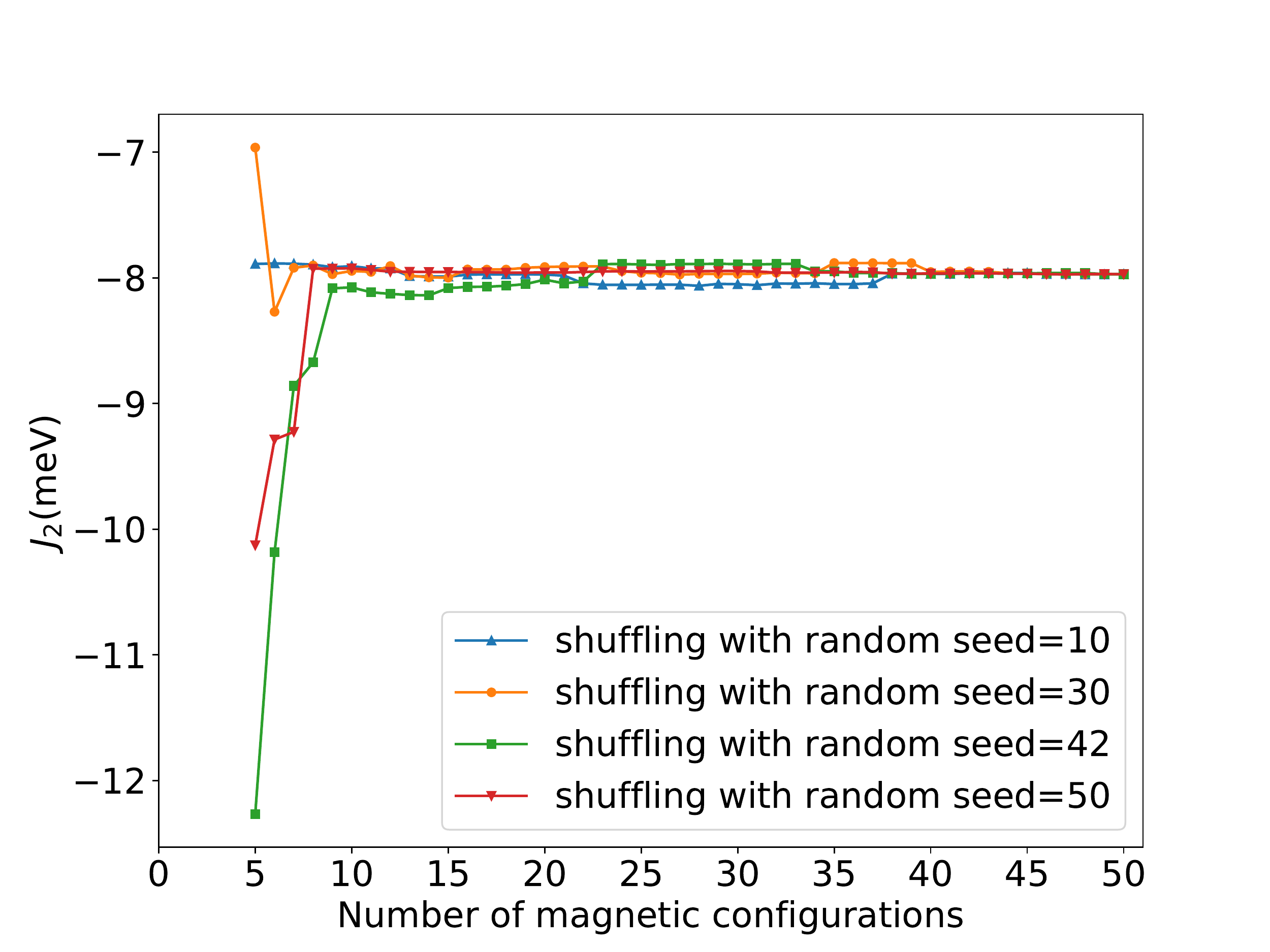}
      \caption {(Color online)  The plots indicate the $J_2$ Heisenberg exchange parameter of MnO (using GGA+$U$ with $U=4$ eV)
                               versus the number of magnetic configurations up to 50.
                               Using minimal configurations (i.e., 5) can result in $\sim50 \%$ error in $J_2$.
                               To indicate that $J_2$ convergence depends on the number of configurations 
                               and not on specific configurations, we plot $J_2$ for different permutations 
                               of the equations related to the 50 magnetic configurations.}
    \label{fig:J2-MnO}
\end{figure}

\begin{figure}
    \centering
      \includegraphics[width=0.95\columnwidth]{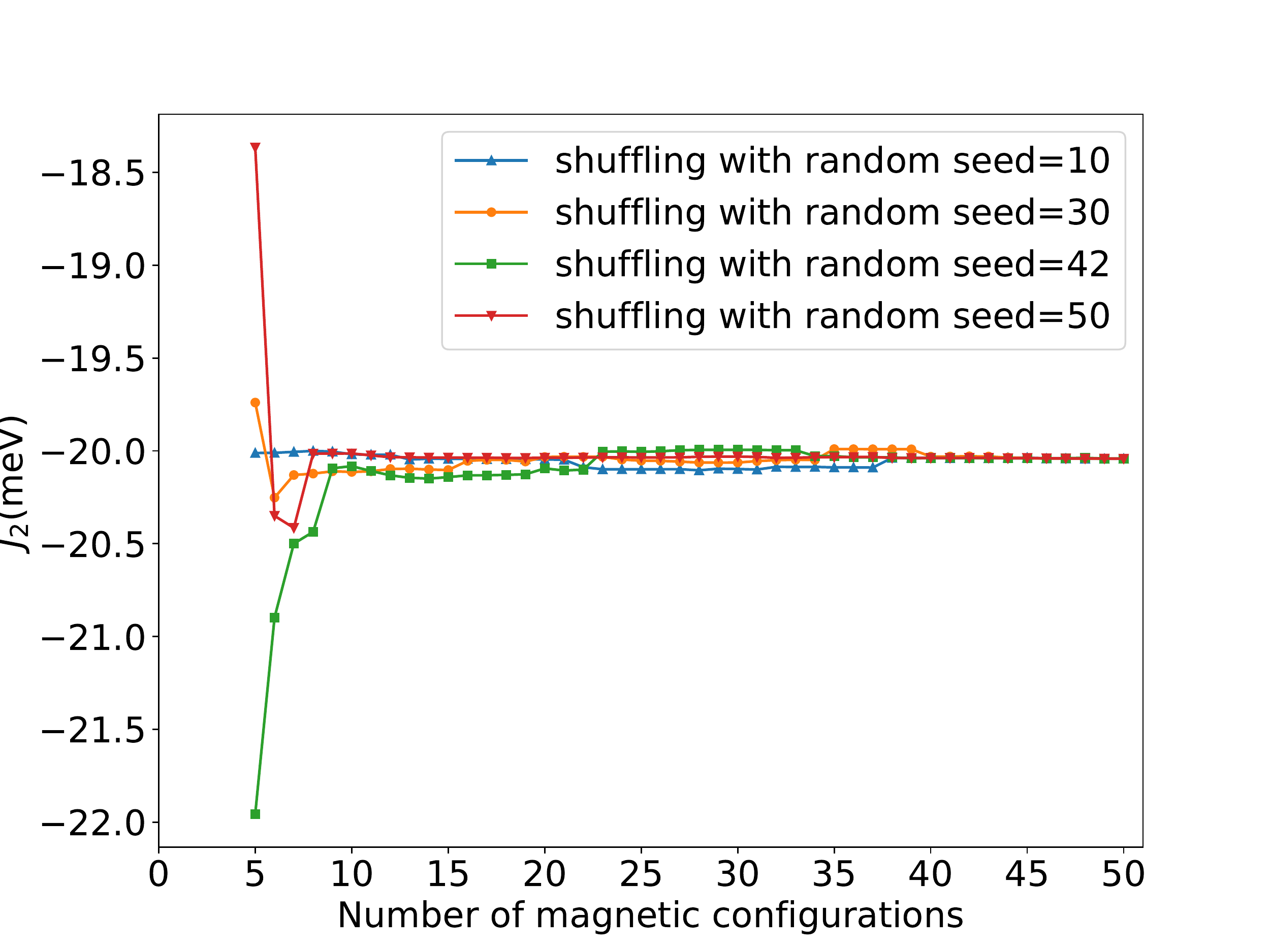}
      \caption {(Color online) The plots indicate the $J_2$ Heisenberg exchange parameter of NiO (using GGA+$U$ with $U=5$ eV) 
                               versus the number of magnetic configurations up to 50. 
                               Using minimal configurations (i.e., 5) can result in $\sim10 \%$ error in $J_2$. 
                               To indicate that $J_2$ convergence depends on the number of configurations 
                               and not on specific configurations, we plot $J_2$ for different permutations 
                               of the equations related to the 50 magnetic configurations.}
    \label{fig:J2-NiO}
\end{figure}

As mentioned, we use a supercell containing 32 atoms to derive exchange Heisenberg interactions.
This supercell allows us to compute $J_{ij}$ up to the fourth nearest neighbor. 
$J_1$, $J_2$, $J_3$, and $J_4$ indicate $J_{ij}$ for the first, second, third, and fourth nearest neighbor. 
The results are presented in tables~\ref{MnO} and ~\ref{NiO} for different $U$ parameters using LDA and GGA.
%
Although only five magnetic configurations are mathematically adequate to obtain these four Heisenberg exchanges, 
using the minimal configurations can produce significant errors in Heisenberg exchanges.
Fig.~\ref{fig:J2-MnO} and ~\ref{fig:J2-NiO} indicate how employing minimal magnetic configurations creates a considerable error in $J_2$ for MnO and NiO. 
We use 50 magnetic configurations (equivalent to 49 linear equations) to predict $J$s through the least-squares procedure in these figures.
We start with 5 magnetic configurations (4 linear equations), gradually add new configurations (i.e., a new linear equation), 
and re-calculate $J$s at each step using the least-squares technique.
%
By increasing the number of magnetic configurations, $J_2$ converges. 
If we shuffle these equations, we can realize that $J_2$ starts from a different value (far from the converged value of $J_2$) for the minimal number of configurations. 
These calculations demonstrate that using minimal magnetic configurations to obtain Heisenberg exchange is not a good approach.
If our model fits DFT results, we expect that adding a new configuration will not change $J_2$ drastically.
But in our Heisenberg spin model, we only consider the fixed magnetic moment of Mn and Ni atoms 
while ignoring that the oxygen atoms have small induced magnetic moments ( $\sim 0-0.2 \mu_B$). 
Further, the magnetic moments of Mn and Ni atoms exhibit tiny fluctuations from one magnetic configuration to another. 
Therefore to average the magnetic moments fluctuation effect (in O, Mn, and Ni atoms), we should consider more magnetic configurations than minimal ones.
According to Fig.~\ref{fig:J2-MnO} and Fig.~\ref{fig:J2-NiO}, a rule of thumb is 
to choose the number of magnetic configurations twice the number of Heisenberg interactions.

To create a clear comprehension of the induced magnetic moments of oxygen,  
we plot the distribution of the normalized magnetic moments of Oxygen atoms 
($\mathrm{M_O}/\max\{\mathrm{M_O}\}$ ) among 50 configurations. 
As the plot indicates, the magnetic moments of the oxygen atoms present seven different values 
($-6/6$, $-4/6$, $-2/6$, $0$, $2/6$, $4/6$, $6/6$). 
We can predict these values by summating the magnetic moments of the oxygen's first nearest neighbors 
({\it i.e.}, summation of Mn magnetic moments, $\mathrm{M_O} \propto \sum_{i=1}^6 \mathrm{M_{Mn_i}}$). 
For example, if three Mn atoms are up and the other three are down spins, 
then $\mathrm{M_O} \approx 0.0 \mu_B$, and if two Mn atoms are up and the other four are down 
$\mathrm{M_O}/\max\{\mathrm{M_O}\} \approx -2/6$. 
Due to such influence of magnetic moments of Mn atoms on O atoms, 
we can not consider the independent magnetic moments for O atoms. 
Hence to account for the effect of the induced magnetism of oxygen atoms,  
instead of implementing them directly in the Heisenberg model, 
we can add more magnetic configurations and take averaging to consider their effect on Heisenberg exchanges.
\begin{figure}
    \centering
      \includegraphics[width=0.95\columnwidth]{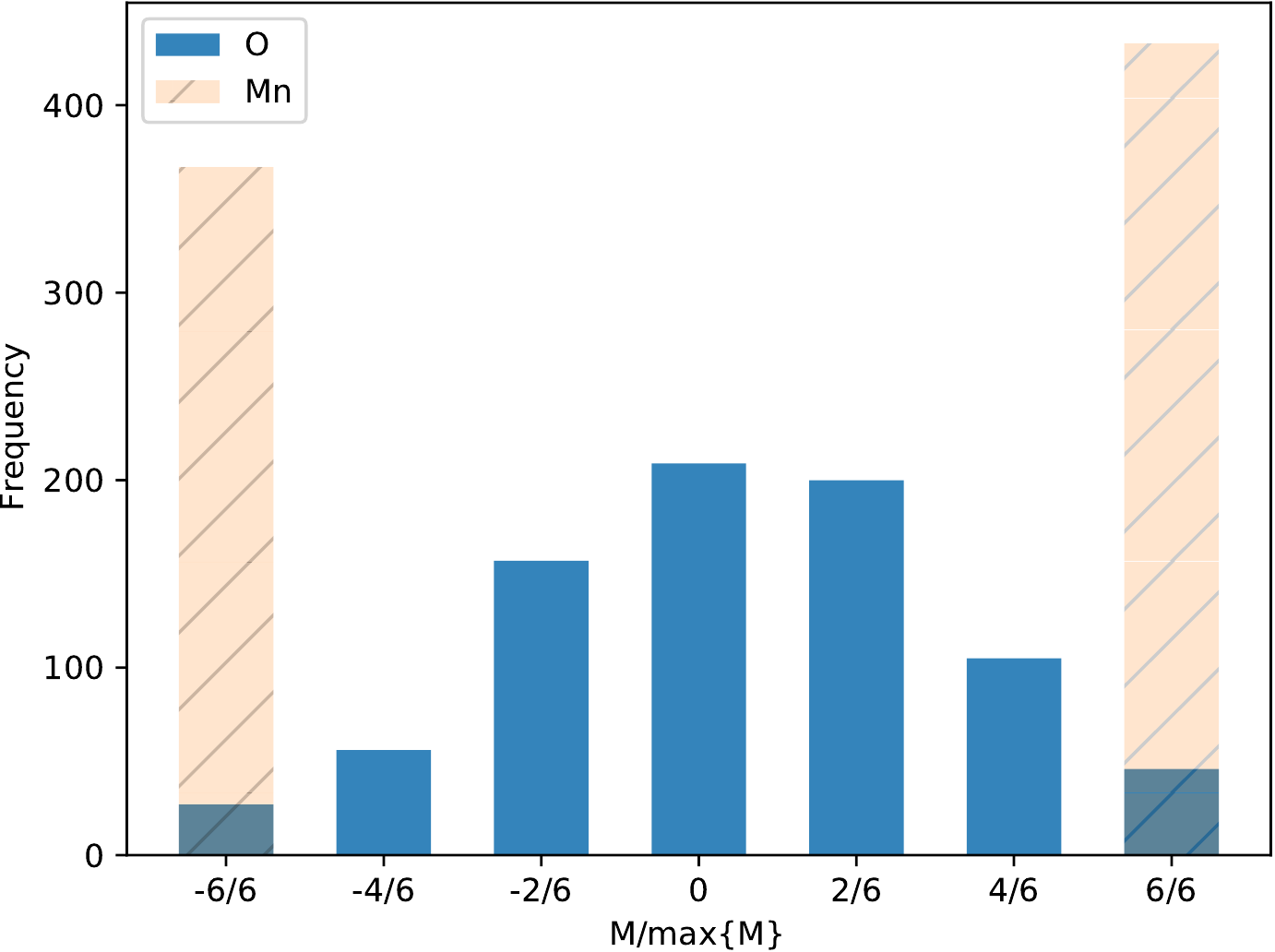}
      \caption {(Color online) The histogram represents the distribution of the normalized magnetic moments, $M/\max\{M\}$, of O and Mn atoms 
                               among 50 magnetic configurations. The Mn atoms indicate fixed magnetizations ($\pm 1$), 
                               while the magnetism of O atoms is induced by the first nearest neighbor Mn magnetic moments. 
                               Due to this effect, the normalized magnetic of O atoms takes some specific values, $\frac{1}{6}\sum_{i=1}^6  \pm 1$, 
                               where 6 is the coordination number of O atoms and $\sum_{i=1}^6  \pm 1$ 
                               is the sum of the normalized magnetic moment on the first nearest neighbor Mn atoms.}
    \label{fig:M_O-dis}
\end{figure}

The third and fourth nearest-neighbor interactions ($J_3$ and $J_4$) 
are negligible compared to $J_2$ and $J_1$ (Table \ref{MnO} and \ref{NiO}) due to electron localization in Mott insulators. 
Hence we only concentrate our analysis on $J_2$ and $J_1$. 
Since LDA prefers more delocalization than GGA, for each value of $U$, LDA+$U$ results in larger Heisenberg exchange values for $J_2$ and $J_1$ than GGA+$U$. 
The proposed $U$ values for MnO and NiO (i.e., $U=6.9$ eV for MnO and $U=8.0$ for NiO) are estimated within the LDA framework. 
For MnO, LDA+$U$ with $U=6.9$ eV predicts $J_2$ and $J_1$, consistent with experimental spin-wave results. 
While for NiO, LDA+$U$ using $U=8.0$ eV underestimates the $J_2$ exchange. 
Generally, $U$ estimation depends on DFT computational details such as muffin-tin radius~\cite{Nawa}. 
Therefore varying the $U$ value around a suggested one can help find more consistent results. 
For example, using $U=7.0$ eV within LDA+$U$ for NiO generates a more consistent result with the neutron scattering experiment for $J_2$ and $J_1$.

\subsection{ Bi-quadratic interaction from DFT calcualtions}
\begin{figure*}
    \centering
      \includegraphics[width=14cm]{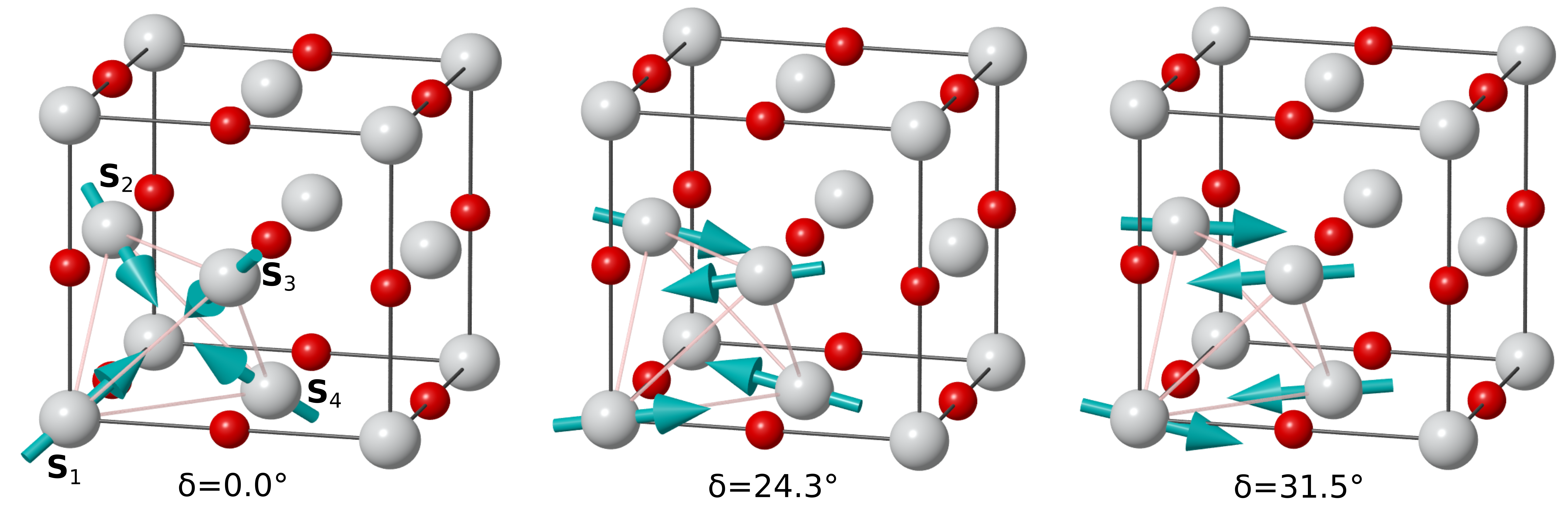}
      \caption {(Color online) The figure illustrates some of the non-collinear magnetic configurations used 
                               to estimate biquadratic interaction. 
                               The direction of magnetic moments ($\mathbf{S}_1$, $\mathbf{S}_2$, $\mathbf{S}_3$, and $\mathbf{S}_4$) 
                               are changed by the $\delta$ parameter according to Eq.~\ref{eq:delta}. }
    \label{fig:mag-configs}
\end{figure*}

We use an approach that we already applied to pyrochlore systems to derive bi-quadratic interaction\cite{fef3-1}. 
As illustrated in Fig.~\ref{fig:mag-configs}, in the cubic cell, there exist four Mn or Ni atoms per cell. 
We start from an all-in magnetic configuration 
(all magnetic moments point to the center of the tetrahedron), in which the summation of magnetic moments is zero. 
The direction of the magnetic moment can be expressed by the polar angle ($\theta$), and azimuthal angle ($\phi$) in the spherical coordinate system.
If we change the direction of these four magnetic moments (in Fig. ~\ref{fig:mag-configs}, we indicate them by $\mathbf{S}_1, \mathbf{S}_2, \mathbf{S}_3$, and $\mathbf{S}_4$) 
by an angle parameter $\delta$ as follows:
\begin{eqnarray} \nonumber
\label{eq:delta}
&&\theta_{1}\rightarrow \theta_{1}+\delta \,\,\,\,\,  \phi_{1} \rightarrow  \phi_{1}-2\delta \\ \nonumber
&&\theta_{2}\rightarrow \theta_{2}-\delta \,\,\,\,\,  \phi_{2} \rightarrow  \phi_{2}+2\delta \\ 
&&\theta_{3}\rightarrow \theta_{3}-\delta \,\,\,\,\,  \phi_{3} \rightarrow  \phi_{3}+2\delta \\ \nonumber 
&&\theta_{4}\rightarrow \theta_{4}+\delta \,\,\,\,\,  \phi_{4} \rightarrow  \phi_{4}-2\delta, \\ \nonumber
\end{eqnarray}
the summation of magnetic moments in the new configuration is still zero ({\it i.e.}, $\mathbf{S}_1+\mathbf{S}_2+\mathbf{S}_3+\mathbf{S}_4=0$).
It can be demonstrated that in such a magnetic configuration, with an arbitrary angle $\delta$, the Heisenberg part of Hamiltonian is degenerate (Appendix \ref{app:a}).
Therefore the DFT energy differences of these magnetic configurations can relate 
to higher powers of $(\mathbf{S}_i \cdot \mathbf{S}_j )$ terms, such as bi-quadratic interactions for the first nearest neighbor.

\begin{figure}
    \centering
      \includegraphics[width=0.85\columnwidth]{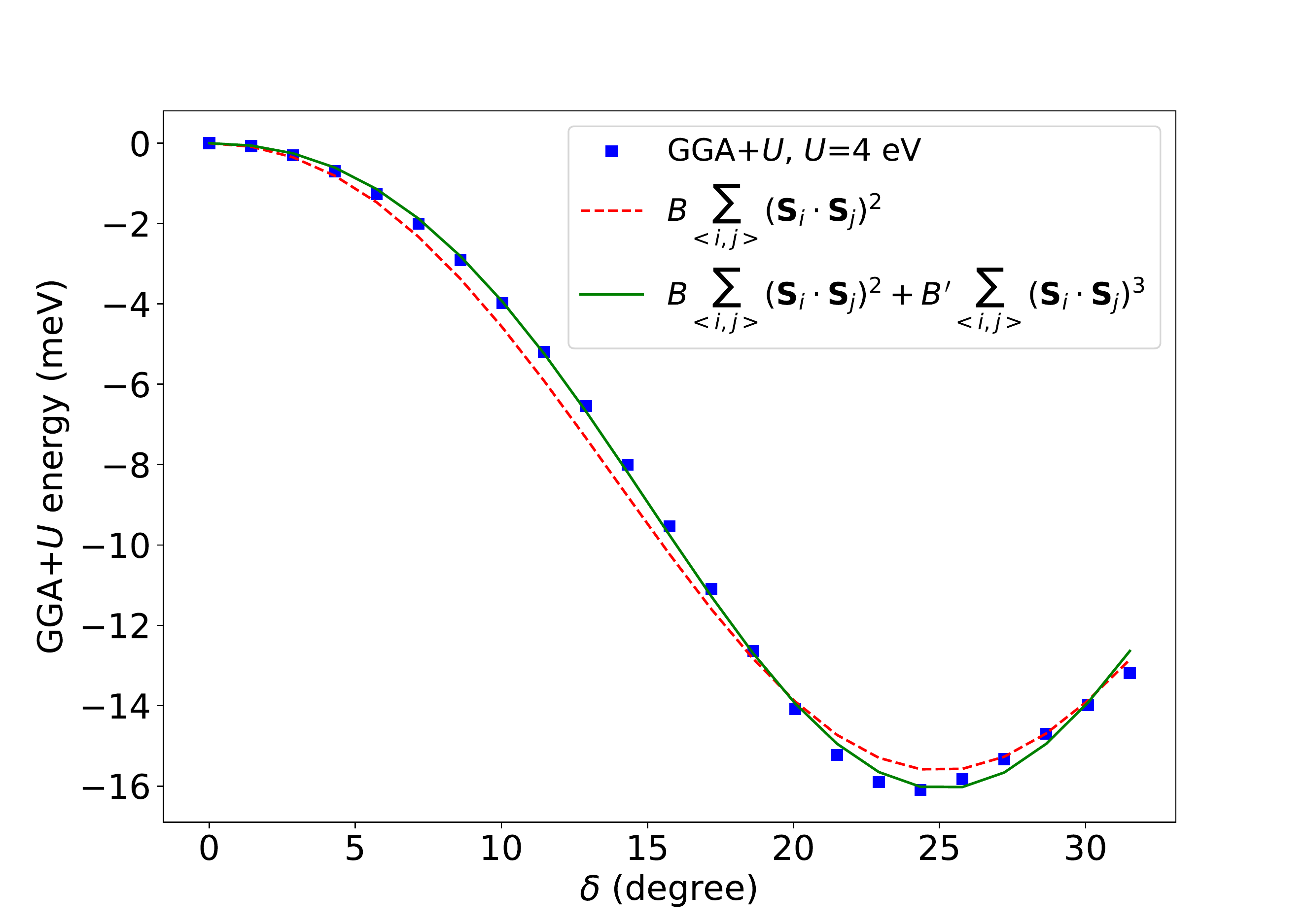}
      \caption {(Color online) The dots data indicate the GGA+$U$ energy of the different non-collinear magnetic configurations used 
                               to calculate the biquadratic parameter ($B$) for MnO. 
                               The data nearly are fitted to biquadratic interaction ($B\sum_{<i,j>} (\mathbf{S}_i \cdot \mathbf{S}_j)^2$). 
                               The data indicate a better fitting to $B\sum_{<i,j>} 
                               (\mathbf{S}_i \cdot \mathbf{S}_j)^2 + B^{\prime}\sum_{<i,j>} (\mathbf{S}_i \cdot \mathbf{S}_j)^3$.
                               In the plot, we set the energy zero to the GGA+U energy of magnetic configuration at  $\delta=0$.}
    \label{fig:bij-mno}
\end{figure}

\begin{figure}
    \centering
      \includegraphics[width=0.85\columnwidth]{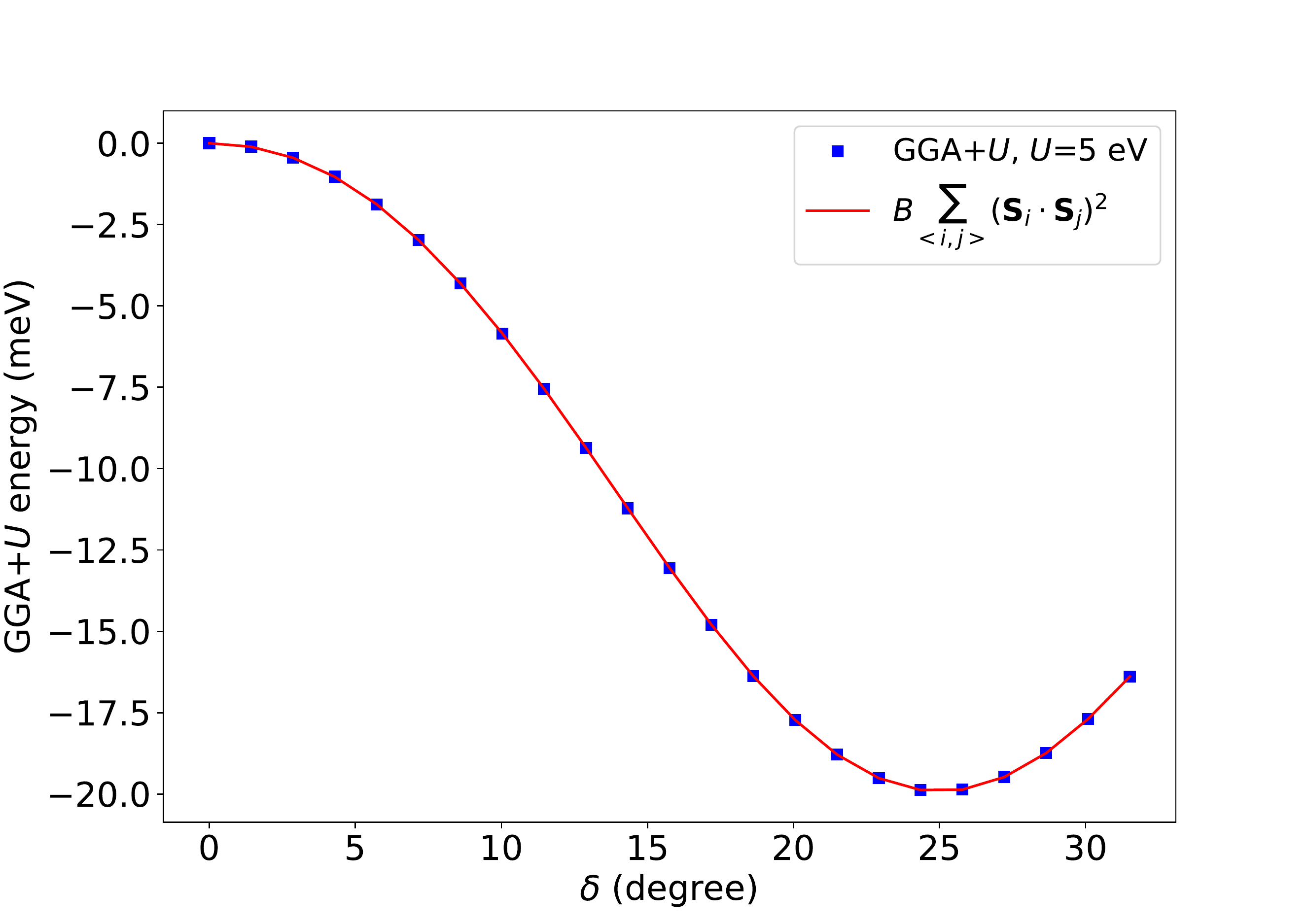}
      \caption {(Color online) The dots data indicate the GGA+$U$ energy of the different non-collinear magnetic configurations used 
                               to calculate the biquadratic parameter ($B$) for NiO. 
                               The data perfectly are fitted to biquadratic interaction ($B\sum_{<i,j>} (\mathbf{S}_i \cdot \mathbf{S}_j)^2$).
                               In the plot, we set the energy zero to the GGA+U energy of magnetic configuration at  $\delta=0$.}
    \label{fig:bij-nio}
\end{figure}

We calculate the DFT energy for different $\delta$, and we find that the bi-quadratic term 
perfectly describes the magnetic interaction beyond the Heisenberg exchange for MnO and NiO.
In Fig.~\ref{fig:bij-nio},  we plot DFT energy versus $\delta$ for NiO using GGA+$U$ with $U=5$ eV.
The figure demonstrates that the DFT energies of these magnetic configurations fit very well to $B \sum_{<i,j>} (\mathbf{\hat{S}}_i \cdot \mathbf{\hat{S}}_j)^2$.
For MnO, fitting the GGA+$U$ data to $B \sum_{<i,j>} (\mathbf{\hat{S}}_i \cdot \mathbf{\hat{S}}_j)^2$  indicates a slight deviation from  DFT results (Fig.~\ref{fig:bij-mno}). 
Essentially, we can expand isotropic coupling up to $(\mathbf{S}_i \cdot \mathbf{S}_j)^{2S}$ for a system with magnetic moments of size $S$ \cite{Fazekas}.
Since for NiO, $S=1$, the bi-quadratic term is sufficient. Whereas  
for MnO, $S=5/2$, therefore $(\mathbf{S}_i \cdot \mathbf{S}_j)^{3}$, $(\mathbf{S}_i \cdot \mathbf{S}_j)^{4}$, and $(\mathbf{S}_i \cdot \mathbf{S}_j)^{5}$ can have contributions.
Our investigation indicates that adding $B^{\prime} \sum_{<i,j>} (\mathbf{\hat{S}}_i \cdot \mathbf{\hat{S}}_j)^{3}$ improves
data fitting (Fig. ~\ref{fig:bij-mno}). 
Adding the remaining higher terms ({\it i.e.}, $(\mathbf{S}_i \cdot \mathbf{S}_j)^{4}$ and $(\mathbf{S}_i \cdot \mathbf{S}_j)^{5}$ ) does not influence the result. 
If we only consider the bi-quadratic term, we obtain $B=-0.83$ meV, and if we consider the additional term, $B^{\prime} \sum_{<i,j>} (\mathbf{\hat{S}}_i \cdot \mathbf{\hat{S}}_j)^{3}$, 
we reach $B=-1.03$ meV and $B^{\prime}=-0.48$ meV. 

The fitting parameter of bi-quadratic interaction strength ($B$) 
is presented for MnO  and NiO within LDA+$U$ and GGA+$U$ for different $U$ parameter values in tables in table~\ref{MnO} and ~\ref{NiO}. 
The value of $B$ within GGA+$U$ is negative for both MnO and NiO, indicating that 
the interaction intends to make the magnetic moments of the first nearest neighbor parallel.  
The LDA+$U$ results for MnO indicate positive values for $B$, which is inconsistent with the experiment~\cite{Rodbell}. 
Indeed, our calculations indicate that LDA for MnO also results in a positive value for $B$; thus, 
this error originates from the LDA defect and can be added to the LDA failure list.
 
\subsection{Monte Carlo simulations}
A proper spin model Hamiltonian can only be considered reliable 
and realistic if the calculated magnetic properties are analogous to the experimental values. 
Motivated by this, we use the obtained spin model Hamiltonian containing  Heisenberg and bi-quadratic terms 
to determine ${T_{\mathrm{N}}}$ and ${\Theta_{\mathrm{CW}}}$) 
by the parallel tempering MC. 
This allocation aims for dual goals of finding the better spin model Hamiltonian 
and the effect of bi-quadratic interaction in thermodynamic magnetic properties of MnO and NiO. 
The results are presented in Tables~\ref{MnO} and~\ref{NiO} for MnO
and NiO, respectively.

\begin{figure}
    \centering
      \includegraphics[width=0.85\columnwidth]{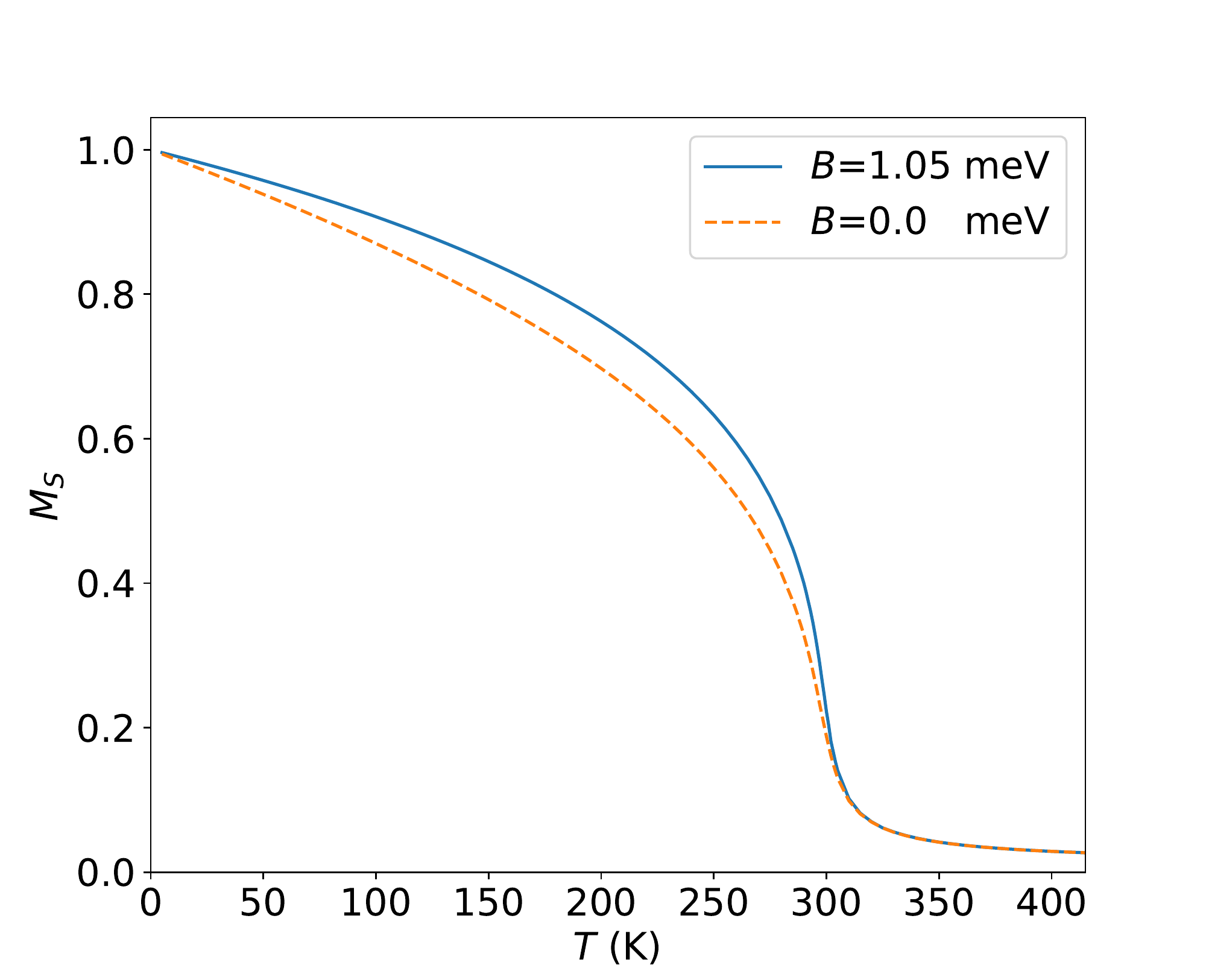}
      \caption {(Color online) The staggered magnetization versus temperature using MC simulations with and without biquadratic interaction.
                               We use exchange parameters obtained from GGA+$U$ with $U=5$ eV for NiO. }
    \label{fig:ms}
\end{figure}

Our MC simulations indicate that including biquadratic interaction has a minimal influence on ${T_{\mathrm{N}}}$. 
For example, the biquadratic term in MC simulation for NiO (using GGA+$U$ with $U = 5$ eV) only causes 
an increasing ${T_{\mathrm{N}}}$ from 295 K to 297 K. However, we can uncover the influence of 
biquadratic interaction through the order parameter. We define the order parameter as the following staggered magnetization:
\begin{equation}
M_s=\frac{1}{N} \Big | \sum_i^{N}  c_i \mathbf{S}_i \Big |, 
\end{equation}
where $\mathbf{S}_i$ is the magnetic moment at $i$th lattice site, $N$ is the total lattice sites, and $c_i=\pm 1$. 
As we illustrate in Fig.~\ref{fig:cubic}, the magnetic order between two successive (111) planes is AF. 
Therefore the sign of $c_i$ changes from one plane to the next nearest plane. 
In Fig.~\ref{fig:ms}, we plot $M_s$ with and without biquadratic interaction.
The figure shows a more sharp variation in $M_s$ in the presence of biquadratic interaction.

Comparing $T_{\mathrm{N}}$ and ${\Theta_{\mathrm{CW}}}$ obtained 
by MC simulations with the experimental values in Tables~\ref{MnO} and~\ref{NiO}, 
it seems the consistent results with the experiment for LDA+$U$ relate to $U \sim 4.0$ eV and $U < 5.0$ eV 
for MnO and NiO, respectively. 
However, at these values of $U,$ the Heisenberg exchange strengths are much stronger than 
experimental exchanges derived from spin-wave theory (Tables~\ref{MnO} and~\ref{NiO}).
In the continue, we shed light on this ambiguity which causes misleading in some research~\cite{Archer}.

If we use spin-wave experimental values for Heisenberg exchanges in the MC simulations, 
the results for $T_{\mathrm{N}}$ are also not satisfactory (Tab.~\ref{Neel-tmp}). 
These seemingly erroneous results are due to our lack of attention to the spin-wave
theory for deriving Heisenberg exchange parameters from neutron scattering experiments. 
The spin-wave approach considers the quantum effect.
Therefore, if we want to use spin-wave exchange parameters 
for classical MC simulations, we should apply $(S+1)/S$ correction~\cite{Fazekas_ch6} (Appendix \ref{app:b}) directly into exchange parameters 
in MC simulations or indirectly in $T_{\mathrm{N}}$ derived from MC simulations.

To demonstrate the necessity of this correction, we consider about ten additional AF
compounds (Tab.~\ref{Neel-tmp}) whose Heisenberg exchange parameters were reported 
by fitting spin-wave theory into neutron scattering data. 
Using the Heisenberg exchange, MC simulation underestimates $T_{\mathrm{N}}$ with a 36\% error on average (Tab.~\ref{Neel-tmp}). 
However, if we multiply $T_{\mathrm{N}}$ by $(S+1)/S$ coefficient, 
then the correction ($T_{\mathrm{N}}^{{\mathrm{MC}}^{*}}= \frac{S+1}{S} T_{\mathrm{N}}^{\mathrm{MC}}$) 
has only an 8\% error on average. 
Therefore, we should include $(S+1)/S$ correction in classical MC simulations 
when considering experimental Heinberg exchanges based on spin-wave analysis.
We conjecture that for DFT+$U,$ also we need such correction. 
In our ongoing research, we are examining this correction for 
Heisenberg exchanges derived from DFT+$U.$ In the present work, 
If we include $(S + 1)/S$ correction for MnO and NiO, 
we can recognize that the estimated $U$ values ({\it i.e.}, $U=6.9$ for MnO and $U=8.0$ for NiO) result in 
more reasonable ${T_{\mathrm{N}}}$ for MnO and NiO.

\begin{table}
\caption{The calculated exchange parameters for MnO using GGA+$U$ and LDA+$U$. 
         The experimental estimation of $J_2$ and $J_1$, using spin-wave,  is also presented. 
         Also, we report the N\'eel ($T_{\mathrm{N}}$) and Curie-Wiess (${\Theta_{\mathrm{CW}}}$) temperatures for each set of exchange parameters obtained by MC simulations. We add the experimental values for $T_{\mathrm{N}}$ and ${\Theta_{\mathrm{CW}}}$ at the end of the table.}\label{MnO}
\centering
\begin{ruledtabular}
\begin{tabular}{rccccccccr}
                     & $J_{1}$    & $J_{2}$    & $J_{3}$     & $J_{4}$    & $B$   & ${T_{\mathrm{N}}}$    & ${\Theta_{\mathrm{CW}}}$  \T \\
\hline                                                                                                                         
GGA                  &            &           &             &            &            &               &                          \TB \\
\hline                                                                                                                         
$U=4.0$   & ${-8.45}$   & ${-7.97}$   & ${-0.27}$    & ${-0.58}$   & ${-0.83}$  & ${82}$  & ${-674}$             \T \\
                                                                                                                               
$U=5.0$   & ${-7.47}$   & ${-6.24}$   & ${-0.21}$    & ${-0.44}$   & ${-0.85}$  & ${62}$  & ${-576}$              \\ 
                                                                                                                               
$U=6.0$   & ${-6.67}$   & ${-4.89}$   & ${-0.16}$    & ${-0.35}$   & ${-0.86}$  & ${46}$  & ${-489}$              \\ 
                                                                                                                               
$U=6.9$   & ${-6.07}$   & ${-3.91}$   & ${-0.14}$    & ${-0.29}$   & ${-0.85}$  & ${34}$  & ${-431}$             \\ 
\hline                                                                                                                         
LDA                  &            &            &             &            &            &            &                          \TB \\
\hline                                                                                                                         
$U=4.0$  & ${-7.41}$   & ${-10.07}$  & ${-0.30}$    & ${-0.71}$   & ${0.30}$   & ${117}$ & ${-696}$             \T \\
                                                                                                                               
$U=5.0$  & ${-6.29}$   & ${-7.97}$   & ${-0.22}$    & ${-0.53}$   & ${0.19}$   & ${92}$  & ${-570}$             \\ 
                                                                                                                               
$U=6.0$  & ${-5.40}$   & ${-6.33}$   & ${-0.17}$    & ${-0.42}$   & ${0.13}$   & ${71}$  & ${-458}$             \\ 
                                                                                                                               
$U=6.9$  & ${-4.76}$   & ${-5.15}$   & ${-0.14}$    & ${-0.34}$   & ${0.09}$   & ${56}$  & ${-386 }$            \\
\hline                                                                                                                         
Exp.                  &            &            &             &            &            &            &                          \TB \\
\hline                                                                                                                         
                      & ${-5.18}$\cite{Pepy1974}   & ${-5.58}\cite{Pepy1974}$   &            &             &         &  ${118}$\cite{coey}    &   ${-610}$\cite{coey}                 \\
                      & ${-4.57}$\cite{Hohlwein}   & ${-5.17}$\cite{Hohlwein}   &            &             &         &           &                 \\
\end{tabular}
\end{ruledtabular}
\end{table}

\begin{table}
\caption{The calculated exchange parameters for NiO using GGA+$U$ and LDA+$U$.
         The experimental estimation of $J_2$ and $J_1$, using spin-wave,  is also presented.
         Also, we report the N\'eel ($T_{\mathrm{N}}$) and Curie-Wiess (${\Theta_{\mathrm{CW}}}$) temperatures 
         for each set of exchange parameters obtained by MC simulations. 
         We add the experimental values for $T_{\mathrm{N}}$ and ${\Theta_{\mathrm{CW}}}$ at the end of the table. }\label{NiO}
\centering
\begin{ruledtabular}
\begin{tabular}{rccccccccr}

                     & $J_{1}$     & $J_{2}$     & $J_{3}$    & $J_{4}$    & $B$     & ${T_{\mathrm{N}}}$      & ${\Theta_{\mathrm{CW}}}$    \T \\
\hline                                                                                    
GGA                  &             &             &            &            &              &              &                            \TB \\
\hline                                                                                    
$U=5.0$   & ${1.66}$   & ${-20.05}$   & ${-0.12}$   & ${-0.82}$   & ${-1.05}$    & ${297}$   & ${-511}$                \T \\
                                                                                                                                      
$U=6.0$   & ${1.23}$   & ${-16.54}$   & ${-0.03}$   & ${-0.67}$   & ${-0.85}$    & ${246}$   & ${-445}$               \\ 
                                                                                                                                      
$U=7.0$   & ${0.98}$   & ${-13.66}$   & ${-0.03}$   & ${-0.51}$   & ${-0.68}$    & ${204}$   & ${-331}$               \\ 
                                                                                                                                      
$U=8.0$   & ${0.78}$   & ${-11.28}$   & ${-0.02}$   & ${-0.40}$   & ${-0.55}$    & ${170}$   & ${-293}$               \\ 
\hline                                                                                                                                      
LDA                  &             &             &            &            &              &              &                            \TB \\
\hline                                                                                                                                      
$U=5.0$   & ${2.28}$   & ${-26.23}$   & ${0.0}$    & ${-1.39}$   & ${-0.96}$    & ${372}$   & ${-688}$               \T \\
                                                                                                                                      
$U=6.0$   & ${1.87}$   & ${-21.65}$   & ${-0.01}$   & ${-1.01}$   & ${-0.78}$    & ${314}$   & ${-560}$               \\ 
                                                                                                                                      
$U=7.0$   & ${1.46}$   & ${-17.79}$   & ${-0.03}$   & ${-0.67}$   & ${-0.63}$    & ${266}$   & ${-454}$               \\ 
                                                                                                                                      
$U=8.0$   & ${1.22}$   & ${-14.77}$   & ${-0.01}$   & ${-0.57}$   & ${-0.50}$    & ${219}$   & ${-367}$               \\ 
\hline                                                                                                                         
Exp.                  &            &            &             &            &            &            &                          \TB \\
\hline                                                                                                                         
                      & ${-1.37}$\cite{Hutchings1972}  &  ${-19.04}$\cite{Hutchings1972}          &            &             &         &  ${525}$\cite{coey}    &   ${-1310}$\cite{coey}                 \\
                      &  ${1.37}$\cite{Shanker}        &  ${-17.32}$\cite{Shanker}                &            &             &         &                        &                                              \\ 
                      &     &  ${-18.1}$\cite{Betto2017}                &            &             &         &                        &                                              \\ 
\end{tabular}
\end{ruledtabular}
\end{table}

\begin{table}[!htb]
\caption{We use MC simulations to obtain N\'eel temperatures ($T_{\mathrm{N}}$) for different compounds. 
We build the Heisenberg spin model for each compound using the exchange interactions 
derived from fitting Neutron scattering data to spin-wave theory.  
By multiplying the MC results to the factor of ${\frac{(S+1)}{S}}$ for each compound, the results, 
($T_{\mathrm{N}}^{{\mathrm{MC}}^{*}}= \frac{S+1}{S} T_{\mathrm{N}}^{\mathrm{MC}}$), 
are in good agreement with the experimental values ($T_{\mathrm{N}}^{\mathrm{exp}}$).} \label{Neel-tmp}
\setlength\tabcolsep{0pt} 
\footnotesize\centering

\smallskip 
\begin{tabular*}{\columnwidth}{@{\extracolsep{\fill}}rcccr}
\hline
Compound                              & ${S}$            & $T_{\mathrm{N}}^{\mathrm{exp}}$  & $T_{\mathrm{N}}^{\mathrm{MC}}$  & $T_{\mathrm{N}}^{{\mathrm{MC}}^{*}}$     \TB \\[1.5ex] 
\hline                                                                                                                
MnO             \cite{Pepy1974}       & ${\frac{5}{2}}$  & 118            & 79   & 110.6           \TB \\[1.5ex]
MnS             \cite{MnS}            & ${\frac{5}{2}}$  & 152            & 106  & 148.4        \\[1.5ex]  
MnTe            \cite{MnTe}           & ${\frac{5}{2}}$  & 310            & 249  & 348          \\[1.5ex] 
MnF$_{2}$       \cite{MnF2}           & ${\frac{5}{2}}$  & 67.46          & 42   & 58.8         \\[1.5ex] 
MnPSe$_{3}$     \cite{MnPSe3}         & ${\frac{5}{2}}$  & 74             & 51   & 71.4         \\[1.5ex] 
FeF$_{2}$       \cite{FeF2}           & ${2}$            & 78.4           & 55   & 82.5         \\[1.5ex] 
LaMnO$_{3}$      \cite{LaMnO3}        & ${2}$            & 139.5          & 98   & 147          \\[1.5ex] 
CrBr$_{3}$      \cite{CrBr3}          & ${\frac{3}{2}}$  & 32.5           & 19   & 31.7         \\[1.5ex]                                                               
Cr$_{2}$O$_{3}$ \cite{Cr2O3}          & ${\frac{3}{2}}$  & 308            & 194  & 323           \\[1.5ex]                                                               
NiBr$_{2}$      \cite{NiBr2}          & ${1}$            & 52             & 29   & 58           \\[1.5ex] 
NiF$_{2}$       \cite{NiF2}           & ${1}$            & 73.2           & 39   & 78           \\[1.5ex] 
NiPS$_{3}$      \cite{NiPS3}          & ${1}$            & 155            & 88   & 176          \\[1.5ex] 
NiO             \cite{Hutchings1972}  & ${1}$            & 525            & 317  & 634           \\[1.5ex] 
\hline
MAE                                   &                  &  -             & 36\% & 8\%         \\
\hline 
\end{tabular*}
\end{table}

\section{conclusion}
We derived  Heisenberg and biquadratic exchanges using GGA+$U$ and LDA+$U$ 
for MnO and NiO and compared the results with experiments through classical MC. 
We indicated that we need to increase the number of magnetic configurations, at least up 
to twice the number of exchange parameters, to obtain more reliable Heisenberg exchanges using DFT. 
To estimate biquadratic exchange, we introduced some non-collinear magnetic configurations 
that can be used for cubic TMOs. We showed that LDA+$U$ predicts wrong biquadratic interaction for MnO due to LDA fault. 
We used estimated exchanges in MC simulations to obtain ${T_{\mathrm{N}}}$ and ${\Theta_{\mathrm{CW}}}$ to evaluate exchange parameters. 
Using the biquadratic exchange in MC simulation has a minimal influence on transition temperature for NiO and MnO. 
However, we indicated that the biquadratic exchange could affect the order parameter. 
We also demonstrated that we should apply $(S+1)/S$ correction in 
the classical MC simulations when using exchange parameters derived from the spin-wave approach. 
We suggest that such correction can also apply to GGA+$U$ and LDA+$U$ results.

\section{Acknowledgments}
This work was supported by the Vice-Chancellor
for Research Affairs of Isfahan University of Technology (IUT).
M. A. thanks Hamid Nouri for help with some Fleur input files.
\begin{appendices}
\section{The degeneracy of the Heisenberg exchange part when $\mathbf{S}_1+\mathbf{S}_2+\mathbf{S}_3+\mathbf{S}_4=0$}
\label{app:a}
If we change the directions of magnetic moments according to equation \ref{eq:delta}
for the cubic structure containing four Mn atoms, $\mathbf{S}_1+\mathbf{S}_2+\mathbf{S}_3+\mathbf{S}_4$ remains constant ($=0$). 
In this case, since:
\begin{eqnarray} \nonumber
(\mathbf{S}_1+\mathbf{S}_2+\mathbf{S}_3+\mathbf{S}_4)^2 &=& \mathbf{S}_1^2+\mathbf{S}_2^2+\mathbf{S}_3^2+\mathbf{S}_4^2 \\ \nonumber
                                                        &+& 2\mathbf{S}_1 \cdot \mathbf{S}_2+ 2\mathbf{S}_1 \cdot \mathbf{S}_3 + 2\mathbf{S}_1 \cdot \mathbf{S}_4 \\  \nonumber
                                                        &+& 2\mathbf{S}_2 \cdot \mathbf{S}_3 + 2\mathbf{S}_2 \cdot \mathbf{S}_4 + 2 \mathbf{S}_3 \cdot \mathbf{S}_4, \\ \nonumber
\end{eqnarray}
then $\sum_{i>j} \mathbf{S}_{i} \cdot \mathbf{S}_{j}$ remains constant: 
\begin{eqnarray} \nonumber
\sum_{i>j}^{4} \mathbf{S}_{i} \cdot \mathbf{S}_{j}  &=& \frac{1}{2} (\mathbf{S}_1+\mathbf{S}_2+\mathbf{S}_3+\mathbf{S}_4)^2 \\ \nonumber
                                                    &-& \frac{1}{2} (\mathbf{S}_1^2+\mathbf{S}_2^2+\mathbf{S}_3^2+\mathbf{S}_4^2)
\end{eqnarray}
Therefore the Heisenberg exchange part is degenerate for these particular configurations.
\section{$(S+1)/S$ correction}
\label{app:b}
According to Heisenberg, the energy between two spins at sites $i$ and $j$ can be expressed by  
$E=-J {\mathbf S}_i \cdot {\mathbf S}_j$ ($J<0$). For  AF arrangement of spins at sites $i$ and $j$, 
the energy in classical physics is $E=JS^2$. In quantum, the energy can be calculated as follows:
\begin{equation}
-J {\mathbf S}_i \cdot {\mathbf S}_j = -\frac{1}{2} J \Big [ ({\mathbf S}_i + {\mathbf S}_j)^2 -  {\mathbf S}_i^2 -{\mathbf S}_j^2  \Big ]. \nonumber
\end{equation}
Since the eigenvalue of ${\mathbf S}_i^2$ and ${\mathbf S}_j^2$ is $S(S +1)$, 
and the smallest value for the energy will be obtained if the state of two spins is a singlet ($|{\mathbf S}_i + {\mathbf S}_j|=0$), then:
\begin{equation}
-J {\mathbf S}_i \cdot {\mathbf S}_j = J (S+1)S .\nonumber
\end{equation}
Therefore,
\begin{equation}
\frac{E_{\mathrm {Quantum}}} {E_{\mathrm {Classic}}} =\frac{S+1}{S}\nonumber
\end{equation}
\end{appendices}
\bibliographystyle{apsrev4-1}
\bibliography{main}
\end{document}